\documentclass{article}
\usepackage{spconf,amsmath,graphicx,hyperref}
\usepackage{cite}
\usepackage{amsmath,amssymb,amsfonts}
\usepackage{algorithmic}
\usepackage{graphicx}
\usepackage{textcomp}
\usepackage{xcolor}
\usepackage{hyperref}
\usepackage{multirow}
\usepackage{booktabs}
\usepackage{circledsteps}
\usepackage{utfsym}
\usepackage{pifont}
\usepackage[table]{xcolor}
\usepackage{threeparttable}

\usepackage{etoolbox}
\makeatletter
\patchcmd{\@maketitle}
  {\@thanks}  
  {\renewcommand{\thefootnote}{\fnsymbol{footnote}}\@thanks\setcounter{footnote}{0}}
  {}{}
\makeatother

\newcommand{\cmark}{\ding{51}} 

\title{SpeakerRPL v2: Robust Open-set Speaker Identification through Enhanced Few-shot Foundation Tuning and Model Fusion}

\renewcommand{\thefootnote}{\fnsymbol{footnote}}

\name{
Zhiyong Chen\textsuperscript{1}\footnotemark[1]\textsuperscript{*}, 
Shuhang Wu\textsuperscript{2}\footnotemark[1]\textsuperscript{*}, 
Yingjie Duan\textsuperscript{1}, 
Xinkang Xu\textsuperscript{1}, 
Xinhui Hu\textsuperscript{1}
}

\address{
\textsuperscript{1}Hithink Research, Zhejiang, China \\
\textsuperscript{2}Shanghai University, Shanghai, China
}

\begin{document}
\ninept
\maketitle
\renewcommand{\thefootnote}{\fnsymbol{footnote}}
\footnotetext[1]{Contributed equally.}
\renewcommand{\thefootnote}{\arabic{footnote}} 
\setcounter{footnote}{0} 
\begin{abstract}
This paper proposes an improved approach for open-set speaker identification based on pretrained speaker foundation models. Building upon the previous Speaker Reciprocal Points Learning framework (V1), we first introduce an enhanced open-set learning objective by integrating reciprocal points learning with logit normalization (LogitNorm) and incorporating adaptive anchor learning to better constrain target speaker representations and improve robustness. Second, we propose a model fusion strategy to stabilize and enhance the few-shot tuning process, effectively reducing result randomness and improving generalization. Furthermore, we introduce a model selection method to ensure optimal performance in model fusion. Experimental evaluations on the VoxCeleb, ESD and 3D-Speaker datasets demonstrate the effectiveness and robustness of the proposed method under diverse conditions. On a newly proposed Vox1-O–like test set, our method reduces the EER from 1.28\% to 0.09\%, achieving a relative reduction of approximately 93\%.
\end{abstract}
\section{Introduction}
\label{sec:intro}
%
Open-set speaker identification is a critical task in speaker recognition. Systems must not only accurately recognize enrolled target speakers \cite{li2023few} but also reliably detect unseen ones \cite{kishan_openfeat_2022}. This capability is particularly crucial in interaction systems integrated with large language models (LLMs) and target speaker recognition frameworks, where precise speaker identification is essential for maintaining coherent and trustworthy interactions. Such applications are becoming increasingly important across diverse scenarios, including interactive agents \cite{xu2025qwen2}. Despite substantial progress enabled by large-scale pretraining~\cite{chen2024eres2netv2, wang-2023}, significant challenges remain—most notably, improving foundation models to make better use of limited enrollment data, ensuring robustness against the high variability of unseen speakers, and developing reliable few-shot tuning strategies that can adapt foundation models to open-set conditions while mitigating the randomness inherent in few-shot training.

Recent advances leveraging pretrained speaker foundation models have shown great promise in addressing these challenges, particularly through efficient adaptation methods such as few-shot learning. Among them, Speaker Reciprocal Points Learning (SpeakerRPL) \cite{chen_adversarial_2021, 11075516, chen2025towards} establishes a strong foundation by explicitly modeling speaker embeddings in a reciprocal manner, thereby enhancing discrimination between target and unknown speakers through enrollment-based few-shot learning. Nevertheless, further improvements are still needed, especially in stabilizing the few-shot adaptation process.

Studies on target speaker voice activity detection (TS-VAD) and speaker diarization emphasize the importance of modeling target speakers with temporal information in dynamic, multi-speaker environments \cite{maeda2025joint, han2025leveraging, 11122260}. Despite differences in tasks, these approaches consistently depend on effective target speaker modeling under limited enrollment data, underscoring the need for more robust target speaker representations in this scenario—a challenge we address in this work through open-set speaker identification.

Motivated by these observations, we aim to advance open-set speaker identification by building upon pretrained speaker foundation models and previous SpeakerRPL (V1) framework. In this paper, we propose an enhanced approach that introduces a refined open-set learning objective by integrating reciprocal points learning with logit normalization (LogitNorm) \cite{wei2022mitigating}, two of the most effective loss functions for open-set learning \cite{zhang2024openood}. This combination imposes stricter constraints on target speaker embeddings and better models the open-set distribution. In addition, we incorporate adaptive anchor learning, which facilitates more flexible speaker representations for unseen speakers.

Recognizing that few-shot tuning often introduces inherent randomness that undermines generalization \cite{wang2020generalizing}, we note that in the SpeakerRPL framework it is highly efficient in terms of both time and computational resources. This efficiency allows us to train a large pool of candidate models for subsequent fusion. To further improve robustness, we propose a model fusion strategy that aggregates complementary models at the score level, thereby mitigating the performance fluctuations inherent to few-shot tuning. In addition, we introduce a novel model selection policy that identifies the most suitable candidate models for fusion, ultimately enhancing robustness and improving open-set speaker identification performance.

\begin{figure*}[ht!]
  \centering
  \includegraphics[width=\linewidth]{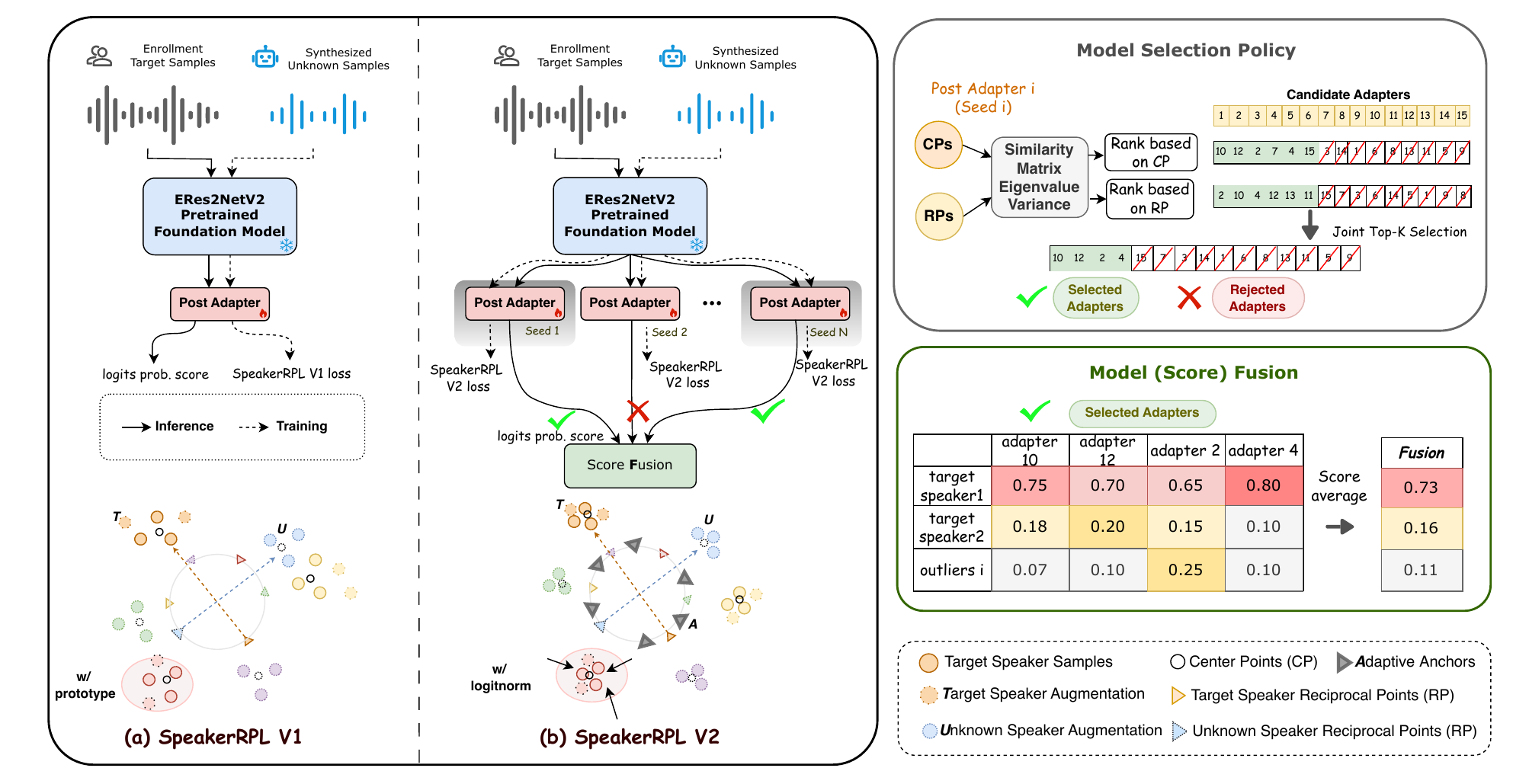}
  \caption{Overall architecture of SpeakerRPL V2.}
  \label{fig:srplv2}
  \vspace{-1em}
\end{figure*}

The primary contributions of this paper are summarized as follows:

\begin{itemize}
\item We propose an improved open-set learning objective integrating reciprocal points learning, logit normalization, and adaptive anchor learning to yield robust speaker embeddings.
\item We introduce an effective model fusion strategy designed to stabilize the few-shot tuning process, thereby reducing inherent randomness and enhancing generalization.
\item We present a post-tuning model selection policy that preserves the most effective tuned models, ensuring optimal fusion performance.
\end{itemize}
Comprehensive experiments of SpeakerRPL V2 on diverse speaker recognition benchmarks validate the effectiveness and reliability of the proposed method across different scenarios.






\section{Methods}
\label{sec:methods}

\subsection{Improving Open-set Speaker Few-shot Learning via Enhanced Reciprocal Points Loss}
To improve the effectiveness of few-shot training and to better focus on the target speaker, we propose integrating logit normalization \cite{wei2022mitigating}, speaker reciprocal points learning \cite{11075516,chen2025towards}, and adaptive anchor normalization. As demonstrated in the left part of Figure~\ref{fig:srplv2}. The loss functions are defined as follows:
\begin{equation}
\resizebox{\linewidth}{!}{$
L_{\text{RPL}} = 
-\log \frac{\exp\!\big(-\mathbf{emb}^\top \mathbf{R}_y\big)}
{\sum_{k=1}^K \exp\!\big(-\mathbf{emb}^\top \mathbf{R}_k\big)}
+ \max\!\left( \|\mathbf{emb} - \mathbf{R}_y\|_2 - \delta, \, 0 \right)
$}
\end{equation}
\begin{align}
z_k &= \mathbf{emb}^\top \mathbf{C}_k, \quad k = 1, \dots, K \\
\tilde{z}_k &= \frac{z_k}{\|\mathbf{z}\|_2}, 
\quad \mathbf{z} = [z_1, \dots, z_K] \\
L_{\text{LogitNorm}} &= -\log 
\frac{\exp(\tilde{z}_y)}
{\sum_{k=1}^K \exp(\tilde{z}_k)}
\end{align}
\begin{equation}
L=L_{LogitNorm} + L_{RPL}
\end{equation}
\noindent
where $\mathbf{emb}$ is the speaker embedding, $\mathbf{R}_k$ and $\mathbf{C}_k$ denote the reciprocal 
point (RP) and classifier center point (CP) of class $k$, and $\delta$ is the learned margin radius. Compared with the 
previous version, the loss $L_{\text{RPL}}$ enforces both discriminability and margin constraints on 
target embeddings, while $L_{\text{LogitNorm}}$ normalizes logits to stabilize classification under 
open-set conditions. The combined objective $L$ jointly enhances robustness and generalization for 
open-set speaker identification. 

Furthermore, we extend the class set $K$ by introducing \emph{adaptive anchors}: 
\[
K = K_{\text{target}} \cup K_{\text{synthetic}} \cup K_{\text{adaptive}}.
\]
Here, $K_{\text{synthetic}}$ corresponds to classes obtained from zero-shot TTS synthesis of unknown 
speakers, while $K_{\text{adaptive}}$ consists of additional reciprocal points that are learned 
dynamically without requiring explicit enrollment data. Notably, the adaptive extension only augments 
$\mathbf{R}_k$ but not $\mathbf{C}_k$, thereby giving the model greater flexibility to allocate representations for unknown speakers in the embedding space. This design enables the system to both strengthen target speaker discrimination and enhance robustness under open-set conditions.

\begin{table*}[h!]
\centering
\caption{Open-set speaker identification results on multiple datasets with 5 target speakers per split (see Table~\ref{tab:label} and Table~\ref{tab:data_info}).}
\resizebox{\linewidth}{!}{
\begin{tabular}{lcccc c cccc c cccc c cccc}
\toprule
\multirow{2}{*}{\textbf{Method}} 
& \multicolumn{4}{c}{\textbf{Settings}} 
& \multicolumn{1}{c}{} 
& \multicolumn{4}{c}{\textbf{VoxCeleb2}} 
& \multicolumn{1}{c}{} 
& \multicolumn{4}{c}{\textbf{3D-Speaker}} 
& \multicolumn{1}{c}{} 
& \multicolumn{4}{c}{\textbf{ESD}} \\

\cmidrule(lr){2-5} \cmidrule(lr){7-10} \cmidrule(lr){12-15} \cmidrule(lr){17-20}
& \textbf{T} & \textbf{U} & \textbf{A} & \textbf{F}
& & EER$\downarrow$ & minDCF$\downarrow$ & OSCR$\uparrow$ & ACC$\uparrow$
& & EER$\downarrow$ & minDCF$\downarrow$ & OSCR$\uparrow$ & ACC$\uparrow$
& & EER$\downarrow$ & minDCF$\downarrow$ & OSCR$\uparrow$ & ACC$\uparrow$ \\

\midrule
\Circled{1} \emph{Direct} Enrollment \cite{chen2024eres2netv2} & - & - & - & - & & 3.74 &0.16 &97.31 &99.21 & & 3.59 & 0.18 & 97.45 & 99.50 & & 4.33 &0.22 &96.17 &98.63\\
\Circled{2} \quad +\textbf{T}arget aug   & \cmark & - &- & - & & 3.84 &0.17 &97.25 &99.24 & & 3.69 & 0.17 & 97.53 & 99.57 & & 4.28 &0.22 &96.27 & 98.71\\
\midrule
\Circled{3} \textbf{SpeakerRPL V1} w/o \textbf{T}arget aug & - & - & -  & - & & 3.76 &0.20 &96.76 &99.42 & & 4.30 &0.23 &95.66 &99.15 & & 5.74 & 0.32 & 92.19 & 98.23\\
\Circled{4} \textbf{SpeakerRPL V1} w/o \textbf{U}nknown speakers  & \cmark & - & - & - & & 3.86 & 0.21 & 96.61 & 99.45 & & 3.39 & 0.19 &96.90 &99.46 & & 5.28 & 0.31 & 93.12 & 98.79\\
\Circled{5} \textbf{SpeakerRPL V1} \cite{chen2025towards} & \cmark & \cmark & - &- & & \cellcolor{yellow!20}0.76 &\cellcolor{yellow!20}0.05 &98.14 & 99.42 & & \cellcolor{yellow!20}0.67 &\cellcolor{yellow!20}0.04 &98.05 &99.38 & & \cellcolor{yellow!20}1.31 & \cellcolor{yellow!20}0.08 & 94.86 & 97.75\\
\midrule
\Circled{6} \textbf{SpeakerRPL V2} w/o fusion & \cmark & \cmark & \cmark & - & & \cellcolor{yellow!20}0.54 & \cellcolor{yellow!20}\textbf{0.03} & 98.04  & 99.42 & & \cellcolor{yellow!20}0.52 &\cellcolor{yellow!20}0.03 &97.89 &99.32 & & \cellcolor{yellow!20}0.94 & \cellcolor{yellow!20}0.06 & 94.93 & 97.71\\
\Circled{7} Softmax \cite{hong2020combining} & \cmark & \cmark & \cmark & - & & 0.69 & 0.04 &97.47 &99.36 & & 0.72 &0.04 &97.09 &99.18 & & 1.47 & 0.08 & 92.26 & 96.02\\
\Circled{8} AM-Softmax \cite{10015805} & \cmark & \cmark & \cmark & - & & 0.58 & 0.04 &97.88 &99.38 & & 0.55 &0.03 &97.79 &99.24 & & 1.04 & 0.06 & 94.23 & 97.15\\
\Circled{9} AAM-Softmax \cite{10094744} & \cmark & \cmark & \cmark & - & & 0.58 & 0.04 &97.85 &99.39 & & 0.58 &0.03 &97.85 &99.29 & & 1.04 & 0.06& 94.00& 97.16\\
\Circled{10} Prototype \cite{li2023few} & \cmark & \cmark & \cmark & - & & 0.72 & 0.05 &93.18 &97.46 & & 1.66 &0.12 &71.88 &82.78 & & 1.67 &0.09 & 91.21 & 95.20\\
\midrule
\Circled{11} \textbf{SpeakerRPL V2} w/ naive fusion & \cmark & \cmark & \cmark & \cmark & & 0.49 &\textbf{0.03} &\cellcolor{yellow!25}98.60 &\textbf{99.51} & & 0.37 & \textbf{0.02} &\cellcolor{yellow!25}98.66 &99.57 & & 0.63 &\textbf{0.04} &\cellcolor{yellow!25}96.60 &98.61\\
\rowcolor{gray!20}
\Circled{12} \textbf{SpeakerRPL V2 (proposed)} & \cmark & \cmark & \cmark & \cmark & & \textbf{0.44} &\textbf{0.03} &\cellcolor{yellow!25}\textbf{98.69} &\underline{99.47} & & \textbf{0.36} &\textbf{0.02} &\cellcolor{yellow!25}\textbf{98.86}&\textbf{99.70} & & \textbf{0.61} & \textbf{0.04} &\cellcolor{yellow!25}\textbf{96.63} & \textbf{98.63}\\
\bottomrule
\end{tabular}
}
\label{tab:exp_main}
\vspace{-1em}
\end{table*}

\begin{table}[h!]
\centering
\caption{Settings used in experiments.}
\begin{tabular}{cl}
\toprule
Symbol & Setting Meaning \\
\midrule
T & \textbf{T}arget speaker augmentation (synthesis) \\
U & \textbf{U}nknown speaker augmentation (synthesis)\\
A & \textbf{A}daptive anchor\\
F & Model \textbf{F}usion\\
\bottomrule
\end{tabular}
\label{tab:label}
\end{table}
\subsection{Robust Few-shot Learning with Model Fusion and Automatic Model Selection Policy}

Few-shot learning often suffers from unstable outcomes, mainly due to variations in parameter initialization and optimizer states. To address this, we employ a model fusion strategy to stabilize training, as demonstrated in Figure~\ref{fig:srplv2}. Specifically, we use score-level averaging, where multiple adapter models are combined to mitigate randomness. For the score output $s_i$ of the $i$-th model candidate trained with random seed $M_i$, the ensembled output is:
\begin{equation}
s = \frac{1}{|\mathcal{S}|} \sum_{i \in \mathcal{S}} s_i, 
\qquad 
\mathcal{S}=\bigl\{\, i \in \{1,\dots,S\} \,\big|\, M_i \text{ is selected} \,\bigr\}.
\end{equation}
A key component is the model selection policy prior to fusion, which identifies effective models among candidates. We observe that well-trained models yield more uniformly distributed central points (CPs) and reciprocal points (RPs). Thus, selection is guided by two criteria: (i) the eigenvalues of the similarity matrix of CPs, and (ii) the eigenvalues of the similarity matrix of RPs. To ensure robustness, only candidates ranking highly on both metrics are preserved.

Let $Z \in \mathbb{R}^{N \times d}$ denote the matrix of $N$ CPs or RPs of dimension $d$ from a single candidate model. The similarity matrix is defined as:
\begin{equation}
C = Z Z^\top, \quad C \in \mathbb{R}^{N \times N}.
\end{equation}
Eigendecomposition yields:
\begin{equation}
C = P \Lambda P^\top, \quad 
\Lambda = \mathrm{diag}(\lambda_1, \lambda_2, \ldots, \lambda_N).
\end{equation}
Let 
\begin{equation}
\boldsymbol{\lambda} = [\lambda_1, \lambda_2, \ldots, \lambda_N]^\top
\end{equation}
denote the eigenvalue vector, whose variance is given by:
\begin{equation}
\mathrm{Var}(\boldsymbol{\lambda}) = \frac{1}{N} \sum_{i=1}^N (\lambda_i - \bar{\lambda})^2, 
\quad \bar{\lambda} = \frac{1}{N}\sum_{i=1}^N \lambda_i.
\end{equation}

The variance of the eigenvalues of the similarity matrix, $\mathrm{Var}(\boldsymbol{\lambda})$, reflects the uniformity of CP and RP distributions. Model candidates with lower variance are preferred, indicating a more balanced embedding space where CPs and RPs are more uniformly distributed (i.e., eigenvalues are closer in magnitude). By leveraging these metrics for ranking with both CPs and RPs, our method automatically selects the most reliable training rounds and their corresponding models, ensuring stable and effective fusion.

\section{Experiments}
\label{sec:exp}

\subsection{Dataset and experimental setting}

The dataset and evaluation details for enrollment (training) and testing is demonstrated in Table~\ref{tab:data_info}.

\begin{table}[!b]
\large
\caption{Detailed information of datasets and evaluation settings.}
\centering
\resizebox{\linewidth}{!}{
\begin{tabular}{lcccc}
\toprule
\multirow{2}{*}{\textbf{Settings}} & \multicolumn{4}{c}{\textbf{Dataset}} \\
\cmidrule(lr){2-5}
& \textbf{VoxCeleb2} & \textbf{3D-Speaker} & \textbf{ESD} & \textbf{Vox1-O*}\\
\midrule
Cross validation splits & 10 & 10 & 5 & 1\\
Enrollment samples & 40 & 30 & 40 & 10 - 88\\
Total speakers & 118 & 73 & 20 & 40\\
Target speakers (per split)  & 5 & 5 & 5 & 40\\
Test known samples (per split)   &1\,653 & 468 & 7\,000 & 3\,369\\
Test unknown samples (per split)   & 34\,569 & 5\,174 & 21\,000 & 36\,237\\
Synthetic unknown speakers (per split) & 50 & 50 & 50 & 40\\
Adaptive anchors (per split) & 50 & 50 & 50 & 40\\
\bottomrule
\end{tabular}
}
\label{tab:data_info}
\end{table}

\begin{table}[h!]
\caption{Experimental Results of Vox1-O* and Vox1-O*-open on 40 target speakers.}
\centering
\resizebox{\linewidth}{!}{
\begin{tabular}{lcccc cccc}
\toprule
\multirow{2}{*}{\textbf{Method}} & \multicolumn{3}{c}{\textbf{Settings}} & \multicolumn{4}{c}{\textbf{Vox1-O*}} \\
\cmidrule(lr){2-4} \cmidrule(lr){5-8}
& \textbf{U} & \textbf{A} & \textbf{F} & EER$\downarrow$ & minDCF$\downarrow$ & OSCR$\uparrow$ & ACC$\uparrow$\\
\midrule
\textit{\textbf{[close-set]}} & & & & & & &\\
\Circled{1} Direct Enrollment \cite{chen2024eres2netv2} & - & - & - &1.28 & 0.07 & 99.76 & 99.76\\
\rowcolor{gray!20}
\Circled{2} \textbf{SpeakerRPL V2} & \cmark & \cmark & \cmark &\textbf{0.09} & \textbf{0.002} & \textbf{99.85} & \textbf{99.85}\\
\midrule
\textit{\textbf{[open-set]}} & & & & & & & \\
\Circled{1} Direct Enrollment \cite{chen2024eres2netv2} & - & - & - &1.72 & 0.08 & 98.02 & 99.76\\
\rowcolor{gray!20}
\Circled{2} \textbf{SpeakerRPL V2} & \cmark & \cmark & \cmark &\textbf{0.24} & \textbf{0.01} & \textbf{99.54} & \textbf{99.85}\\
\bottomrule
\end{tabular}
}
\label{tab:vox1}
\end{table}

\begin{itemize}
\item \textbf{VoxCeleb2:} Sourced from Youtube, this dataset \cite{nagrani2017voxceleb} includes a broad range of real-world samples of various speakers. The training part is following the Voxwatch \cite{peri_voxwatch_2023} settings for focus on target speaker group. The target-speaker watch list is setting to 5 (and also for all other datasets, shown in Table~\ref{sec:exp}).
\item \textbf{3D-Speaker:} A large-scale multi-condition speaker corpus. It includes over 10,000 speakers and more than 579,000 utterances across multiple devices, distances, and dialects \cite{zheng20233d}. We select near-field Mandarin samples for enrollment, far-field samples from different dialects for testing.
\item \textbf{ESD:} It contains 10 English and 10 Mandarin speakers, with  350 utterances for each emotion per speaker across 5 distinct emotional settings \cite{zhou2021seen}. The data is relatively clean for each speaker. Enrollment samples are all selected from the neutral state and testing ones from 4 emotional states (happiness, anger, sadness and surprise).
\item \textbf{Vox1-O*:} Based on the VoxCeleb1-O \cite{nagrani2017voxceleb} testing trials, adapted to the open-set speaker identification protocol. We partially select trials to ensure that the enrollment set (left column) does not overlap with the testing set (right column). In addition, VoxCeleb2 test samples are added to the right column to construct an extended open-set version, Vox1-O*-open.
\end{itemize}


We evaluate our method on both multi–target speaker open-set identification and verification benchmarks with a larger number of speakers, such as VoxCeleb-O. Performance is assessed using EER, minDCF, target speaker closed-set accuracy (ACC), and the open-set recognition rate (OSCR) \cite{chen_adversarial_2021, 11075516}. Details of the experimental settings are demonstrated in Table~\ref{tab:label} and Table~\ref{tab:data_info}. Synthetic speech samples are generated with GPT-SoVITSv2~\cite{rvc-boss-no-date}, while timbres of unknown speakers are selected from LibriTTS and AiShell \cite{zen2019libritts, fu2021aishell}. The adapters are lightweight MLPs. Both the implementation and datasets are publicly available in our repository\footnote{\url{https://github.com/zhiyongchenGREAT/Few-shot-Robust-Speaker-TTS/tree/v2.1}}. The fine-tuning process completes within a few minutes on GPU. For experiments without fusion, we report the average of the metrics over 10 training seeds. For fusion experiments, we train 30 adapter candidates and apply the proposed model selection policy. Specifically, we discard the bottom 33\% of candidates based on both RP- and CP-based eigenvalue variances and retain their intersection, which contains at least 33\% of the models. Other implementation details follow SpeakerRPL V1 \cite{chen2025towards} and are available in the open-source resources.

\begin{table}[h]
\large
\caption{Ablation study on the number of adaptive anchors.}
\centering
\resizebox{0.7\linewidth}{!}{
\begin{tabular}{lccccc}
\toprule
\multirow{2}{*}{\textbf{Metrics}} & \multicolumn{5}{c}{\textbf{Adaptive Anchor Number}} \\
\cmidrule(lr){2-6}
& 10 & 20 & 30 & 40 &50\\
\midrule
EER(\%)$\downarrow$ &\cellcolor{blue!2}0.60 &\cellcolor{blue!5}0.56 &\cellcolor{blue!8}0.53 &\cellcolor{blue!12}0.45 &\cellcolor{blue!18}\textbf{0.42}\\
minDCF(\%)$\downarrow$ &\cellcolor{blue!2}0.04 &\cellcolor{blue!5}0.04 &\cellcolor{blue!8}0.04 &\cellcolor{blue!12}0.03 &\cellcolor{blue!18}\textbf{0.03}\\
OSCR(\%)$\uparrow$ &\cellcolor{blue!2}97.47 &\cellcolor{blue!5}97.30 &\cellcolor{blue!8}97.66 &\cellcolor{blue!12}97.83 &\cellcolor{blue!18}\textbf{98.13}\\
ACC(\%)$\uparrow$ &\cellcolor{blue!2}98.80 &\cellcolor{blue!5}98.66 &\cellcolor{blue!8}98.97 &\cellcolor{blue!12}98.97 &\cellcolor{blue!18}\textbf{99.00}\\

\bottomrule
\end{tabular}
}
\label{tab:anchor_num}
\end{table}

\begin{figure}[!h]
  \centering
  \includegraphics[width=\linewidth]{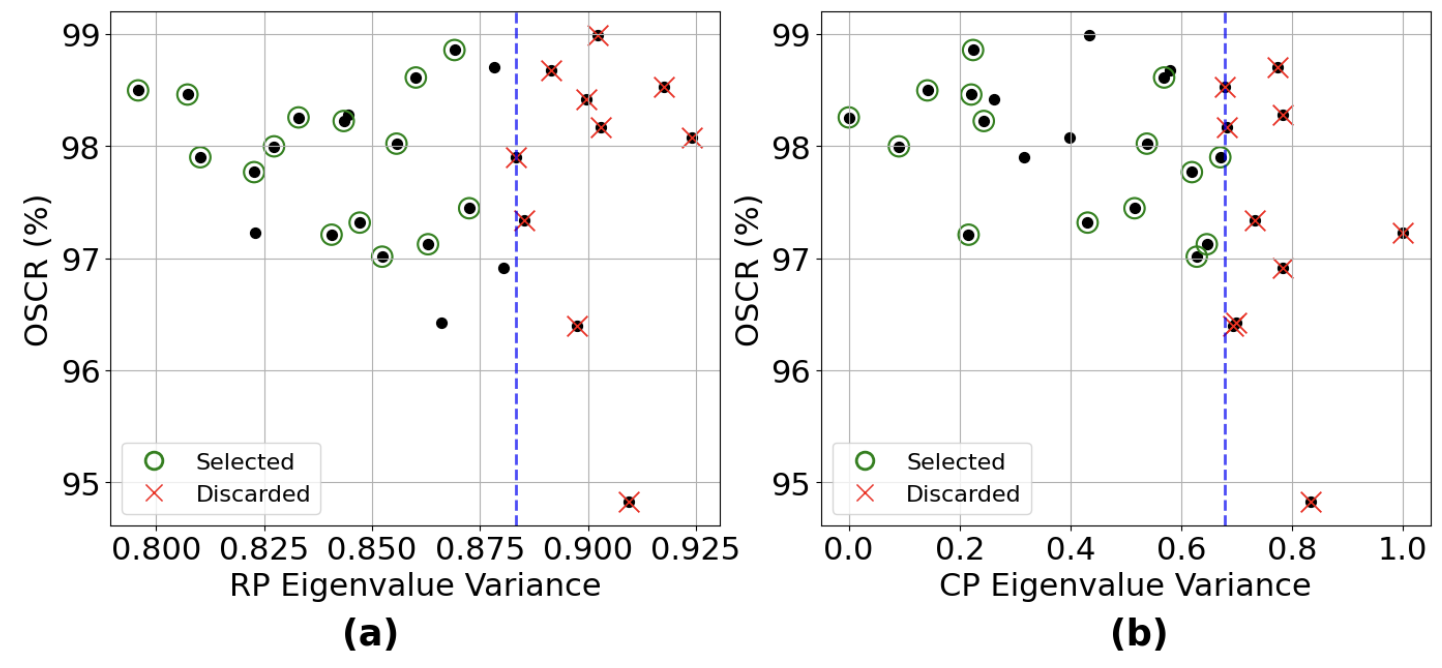}
\caption{Visualization of model selection for fusion based on eigenvalue variance of Reciprocal Points (a) and Center Points (b), and their correlation with candidate models' open-set recognition performance.}
  \label{fig:eigen_var}
  \vspace{-1em}
\end{figure}

\subsection{Results}

We first evaluate the proposed method on multiple open-set speaker identification datasets, as shown in Table~\ref{tab:exp_main}. With direct enrollment, the model achieves baseline performance across all datasets (\Circled{1}), which is the baseline approach taken in wide range of speaker recognition and relevant applications. When applying target speaker augmentation with synthetic data (\Circled{2}) directly, the results are comparable to those obtained with direct enrollment. We then examine the enrollment-time tuning strategy, SpeakerRPL V1 (\Circled{3}–\Circled{5}), which incorporates both target speaker augmentation and unknown speaker augmentation using TTS, following the same settings as the original SpeakerRPL V1 \cite{chen2025towards}. Notably, combining target and unknown speaker augmentation yields the significantly better performance.

With the adaptive anchor mechanism and the improved loss design of SpeakerRPL V2 (\Circled{6}), the model achieves further improvements. In particular, adaptive anchors provide additional modeling flexibility, leading to better representations of both target and unknown speakers. We observe that combining synthetic unknown samples with adaptive anchors yields significant performance gains (\Circled{6}), especially in terms of EER and minDCF (comparing \Circled{5} and \Circled{6}). Building on this result, we conduct ablation studies with different comparable loss functions (\Circled{7}–\Circled{10}), which demonstrate that the proposed combination of LogitNorm and reciprocal points learning in SpeakerRPL V2 performs best. Notably, all the above evaluations are conducted with 10 random seed runs for each split. Since the scores vary across runs, we report the average of the evaluation metrics. 

When applying the proposed SpeakerRPL V2 with model fusion, we observe substantial gains in EER, minDCF, and ACC, as well as consistent improvements in the multiclass metric OSCR—an area where steady improvements have been difficult to achieve with previous baselines. A key finding in (\Circled{11}–\Circled{12}) is that the proposed fusion strategy further enhances performance, with the primary improvement reflected in the stabilization of OSCR gains. While naive score-level averaging of 30 models trained with different seeds yields improvements (\Circled{11}), it also includes poorly performing candidates. By contrast, the proposed model selection policy based on eigenvalue variance (\Circled{12}) produces more robust and rigorous results across all metrics.

As shown in Table~\ref{tab:vox1}, to further extend the findings to broader applications and evaluations, we also assess the method in scenarios with a larger number of target speakers enrolled simultaneously. Specifically, we slightly modify the standard Vox1-O trials to construct Vox1-O* and its corresponding open-set setting, which includes 40 target speakers and ensures that the enrollment subset (left) does not overlap with the testing subset (right). In this configuration, the left subset is used for enrollment and the right subset for testing, which aligns with the open-set speaker identification protocol, while still allowing computation and comparison of metrics with the selected trials in verification settings. Furthermore, we incorporate the VoxCeleb2 test set into the Vox1-O* trials (added to the right testing subset) to form an extended open-set version, Vox1-O*-open. Across these settings, performance improvements remain consistent, demonstrating the robustness of the proposed SpeakerRPL V2.

We analyze the effect of varying the number of adaptive anchors in SpeakerRPL V2. The results in Table~\ref{tab:anchor_num} show that increasing the number of anchors improves performance, with gains saturating at around 50 anchors. This configuration is adopted across all experimental settings.

Finally, we investigate the effectiveness of the proposed model selection policy in Fig.~\ref{fig:eigen_var}. We evaluate the relationship between OSCR performance and the variance of eigenvalues of RPs and CPs to assess the effectiveness of these metrics. The results show a clear distinction between preserved and discarded models under the proposed policy. We observe that models with lower eigenvalue variance consistently achieve higher recognition performance, highlighting the strong correlation between these metrics and model effectiveness. With this policy, lower-performing models are effectively filtered out, while higher-performing models are retained.

\vspace{-1em}

\section{Conclusion}
\label{sec:majhead}

Open-set speaker identification is a critical task within the broader field of speaker recognition. Although enrollment-time few-shot tuning achieves competitive performance, further improvements are needed to enhance robustness and stability across metrics. In this work, we propose three key enhancements. First, we integrate LogitNorm with the SpeakerRPL loss, in combination with adaptive anchor learning, to improve speaker representation. Second, we introduce a model fusion strategy to stabilize the few-shot training process. Third, we enhance the fusion process with a model selection policy based on eigenvalue variance analysis of reciprocal and center points. Experimental results demonstrate that our approach yields robust performance across diverse scenarios of open-set identification, including settings with a larger number of enrolled speakers and evaluation trials resembling speaker verification. In future work, we plan to extend this framework to diarization and other related tasks.

\vfill\pagebreak




\bibliographystyle{IEEEbib}
\bibliography{strings,refs}

\begin{thebibliography}{10}

\bibitem{li2023few}
Yanxiong Li, Hao Chen, Wenchang Cao, Qisheng Huang, and Qianhua He,
\newblock ``Few-shot speaker identification using lightweight prototypical network with feature grouping and interaction,''
\newblock {\em IEEE Transactions on Multimedia}, 2023.

\bibitem{kishan_openfeat_2022}
K~C Kishan, Zhenning Tan, Long Chen, Minho Jin, Eunjung Han, Andreas Stolcke, and Chul Lee,
\newblock ``{OpenFEAT}: {Improving} {Speaker} {Identification} by {Open}-{Set} {Few}-{Shot} {Embedding} {Adaptation} with {Transformer},''
\newblock in {\em {ICASSP} 2022 - 2022 {IEEE} {International} {Conference} on {Acoustics}, {Speech} and {Signal} {Processing} ({ICASSP})}, 2022, pp. 7062--7066,
\newblock ISSN: 2379-190X.

\bibitem{xu2025qwen2}
Jin Xu, Zhifang Guo, Jinzheng He, Hangrui Hu, Ting He, Shuai Bai, Keqin Chen, Jialin Wang, Yang Fan, Kai Dang, et~al.,
\newblock ``Qwen2.5-omni technical report,''
\newblock {\em arXiv preprint arXiv:2503.20215}, 2025.

\bibitem{chen2024eres2netv2}
Yafeng Chen, Siqi Zheng, Hui Wang, Luyao Cheng, Qian Chen, Shiliang Zhang, and Junjie Li,
\newblock ``Eres2netv2: Boosting short-duration speaker verification performance with computational efficiency,''
\newblock in {\em Proc. Interspeech 2024}, 2024, pp. 3245--3249.

\bibitem{wang-2023}
Hui Wang, Siqi Zheng, Yafeng Chen, Luyao Cheng, and Qian Chen,
\newblock ``{CAM++: a fast and efficient network for speaker verification using Context-Aware Masking},''
\newblock {\em Interspeech 2022}, 8 2023.

\bibitem{chen_adversarial_2021}
Guangyao Chen, Peixi Peng, Xiangqian Wang, and Yonghong Tian,
\newblock ``Adversarial {Reciprocal} {Points} {Learning} for {Open} {Set} {Recognition},''
\newblock {\em IEEE Transactions on Pattern Analysis and Machine Intelligence}, pp. 1--1, 2021,
\newblock arXiv: 2103.00953.

\bibitem{11075516}
Zhiyong Chen, Shuhang Wu, Xinnuo Li, Zhiqi Ai, and Shugong Xu,
\newblock ``Open-set speaker identification through efficient few-shot tuning with speaker reciprocal points and unknown samples,''
\newblock {\em IEEE Transactions on Audio, Speech and Language Processing}, vol. 33, pp. 3347--3362, 2025.

\bibitem{chen2025towards}
Zhiyong Chen, Shuhang Wu, Xinnuo Li, Zhiqi Ai, and Shugong Xu,
\newblock ``Towards robust speaker recognition against intrinsic variation with foundation model few-shot tuning and effective speech synthesis,''
\newblock in {\em Proc. Interspeech 2025}, 2025, pp. 1118--1122.

\bibitem{maeda2025joint}
Chikara Maeda, Muhammad Shakeel, and Yui Sudo,
\newblock ``Joint target-speaker asr and activity detection,''
\newblock in {\em Proc. Interspeech 2025}, 2025, pp. 1683--1687.

\bibitem{han2025leveraging}
Jiangyu Han, Federico Landini, Johan Rohdin, Anna Silnova, Mireia Diez, and Luk{\'a}{\v{s}} Burget,
\newblock ``Leveraging self-supervised learning for speaker diarization,''
\newblock in {\em ICASSP 2025-2025 IEEE International Conference on Acoustics, Speech and Signal Processing (ICASSP)}. IEEE, 2025, pp. 1--5.

\bibitem{11122260}
Ming Cheng and Ming Li,
\newblock ``Multi-input multi-output target-speaker voice activity detection for unified, flexible, and robust audio-visual speaker diarization,''
\newblock {\em IEEE Transactions on Audio, Speech and Language Processing}, vol. 33, pp. 3522--3536, 2025.

\bibitem{wei2022mitigating}
Hongxin Wei, Renchunzi Xie, Hao Cheng, Lei Feng, Bo~An, and Yixuan Li,
\newblock ``Mitigating neural network overconfidence with logit normalization,''
\newblock in {\em International conference on machine learning}. PMLR, 2022, pp. 23631--23644.

\bibitem{zhang2024openood}
Jingyang Zhang, Jingkang Yang, Pengyun Wang, Haoqi Wang, Yueqian Lin, Haoran Zhang, Yiyou Sun, Xuefeng Du, Kaiyang Zhou, Wayne Zhang, Yixuan Li, Ziwei Liu, Yiran Chen, and Hai Li,
\newblock ``Open{OOD} v1.5: Enhanced benchmark for out-of-distribution detection,''
\newblock in {\em NeurIPS 2023 Workshop on Distribution Shifts: New Frontiers with Foundation Models}, 2024.

\bibitem{wang2020generalizing}
Yaqing Wang, Quanming Yao, James~T Kwok, and Lionel~M Ni,
\newblock ``Generalizing from a few examples: A survey on few-shot learning,''
\newblock {\em ACM computing surveys (csur)}, vol. 53, no. 3, pp. 1--34, 2020.

\bibitem{hong2020combining}
Qian-Bei Hong, Chung-Hsien Wu, Hsin-Min Wang, and Chien-Lin Huang,
\newblock ``Combining deep embeddings of acoustic and articulatory features for speaker identification,''
\newblock in {\em ICASSP 2020-2020 IEEE International Conference on Acoustics, Speech and Signal Processing (ICASSP)}. IEEE, 2020, pp. 7589--7593.

\bibitem{10015805}
Weiwei Lin and Man-Wai Mak,
\newblock ``Robust speaker verification using deep weight space ensemble,''
\newblock {\em IEEE/ACM Transactions on Audio, Speech, and Language Processing}, vol. 31, pp. 802--812, 2023.

\bibitem{10094744}
Leying Zhang, Zhengyang Chen, and Yanmin Qian,
\newblock ``Adaptive large margin fine-tuning for robust speaker verification,''
\newblock in {\em ICASSP 2023 - 2023 IEEE International Conference on Acoustics, Speech and Signal Processing (ICASSP)}, 2023, pp. 1--5.

\bibitem{nagrani2017voxceleb}
Arsha Nagrani, Joon~Son Chung, and Andrew Zisserman,
\newblock ``Voxceleb: A large-scale speaker identification dataset,''
\newblock {\em Interspeech 2017}, p. 2616, 2017.

\bibitem{peri_voxwatch_2023}
Raghuveer Peri, Seyed~Omid Sadjadi, and Daniel Garcia-Romero,
\newblock ``{VoxWatch}: {An} open-set speaker recognition benchmark on {VoxCeleb},'' June 2023,
\newblock arXiv:2307.00169 [cs, eess].

\bibitem{zheng20233d}
Siqi Zheng, Luyao Cheng, Yafeng Chen, Hui Wang, and Qian Chen,
\newblock ``3d-speaker: A large-scale multi-device, multi-distance, and multi-dialect corpus for speech representation disentanglement,''
\newblock {\em CoRR}, 2023.

\bibitem{zhou2021seen}
Kun Zhou, Berrak Sisman, Rui Liu, and Haizhou Li,
\newblock ``Seen and unseen emotional style transfer for voice conversion with a new emotional speech dataset,''
\newblock in {\em ICASSP 2021-2021 IEEE International Conference on Acoustics, Speech and Signal Processing (ICASSP)}. IEEE, 2021, pp. 920--924.

\bibitem{rvc-boss-no-date}
Rvc-Boss,
\newblock ``{GitHub - RVC-Boss/GPT-SoVITS: 1 min voice data can also be used to train a good TTS model! (few shot voice cloning)},'' .

\bibitem{zen2019libritts}
Heiga Zen, Viet Dang, Rob Clark, Yu~Zhang, Ron~J Weiss, Ye~Jia, Zhifeng Chen, and Yonghui Wu,
\newblock ``Libritts: A corpus derived from librispeech for text-to-speech,''
\newblock in {\em Proc. Interspeech 2019}, 2019, pp. 1526--1530.

\bibitem{fu2021aishell}
Yihui Fu, Luyao Cheng, Shubo Lv, Yukai Jv, Yuxiang Kong, Zhuo Chen, Yanxin Hu, Lei Xie, Jian Wu, Hui Bu, et~al.,
\newblock ``Aishell-4: An open source dataset for speech enhancement, separation, recognition and speaker diarization in conference scenario,''
\newblock in {\em Proc. Interspeech 2021}, 2021, pp. 3665--3669.

\end{thebibliography}

\end{document}